\documentclass[12pt]{article}
\usepackage{epsf}
\usepackage{color}
\usepackage{amsmath,amssymb,color,graphics,amscd,amsfonts}

\newcommand{\be}{\begin{equation}}
\newcommand{\ee}{\end{equation}}
\newcommand{\bea}{\begin{eqnarray}\displaystyle}
\newcommand{\eea}{\end{eqnarray}}
\newcommand{\nn}{\nonumber}

\def\a{\alpha}
\def\b{\beta}

\def\g{\gamma}

\def\e{\epsilon}

\def\s{\sigma}

\def\C{\mathbb{C}}

\def\cO{\mathcal{O}} 
\def\cM{\mathcal{M}} 
\def\tv{ \tilde{v} } 
\def\tw{ \tilde{w} } 

\begin{document}

{}~
{}~
\hbox{QMUL-PH-08-03}
\break

\vskip .6cm

\centerline{{\LARGE \bf  Holomorphic maps and  the complete
 }} 
\centerline{{\LARGE \bf  ${ 1\over N} $ expansion of 2D  $SU(N)$ Yang-Mills  }}

\medskip

\vspace*{4.0ex}

\centerline{ {\large \bf Yusuke Kimura}\footnote{y.kimura@qmul.ac.uk}
{ \bf  and }{\large \bf Sanjaye Ramgoolam}\footnote{s.ramgoolam@qmul.ac.uk}  } 
\vspace*{4.0ex}
\begin{center}
{\large Centre for Research in String Theory, \\
 Department of Physics, \\
Queen Mary, University of London\\
Mile End Road\\
London E1 4NS UK\\
}
\end{center}

\vspace*{5.0ex}

\centerline{\bf Abstract} \bigskip

We give a description of the complete $1/N$ expansion of 
$SU(N)$  2D Yang Mills theory  in terms of the moduli space of
holomorphic maps from non-singular worldsheets.  
This is related to the Gross-Taylor coupled $1/N$ expansion
through a map from Brauer algebras to symmetric groups. 
These results point to an equality between Euler characters 
of moduli spaces of holomorphic maps from non-singular worldsheets with  
a target Riemann surface equipped with markings on the one hand 
and Euler characters of another moduli space involving 
worldsheets with double points  (nodes).

\thispagestyle{empty}
\vfill

\eject

%%%%%%%%%%%%%%%%%%%%%%%%%%%%%%%%%%%%%%%%%%%%%%%%%%%%%%%%%%5

%%%%%%%%%%%%%%%%%%%%%%%%%%%%%%%%%%%%%%%

\section{Introduction} 

The exact partition function \cite{mig}  of two dimensional Yang Mills
(2DYM) for $SU(N)$ gauge group on a Riemann surface of 
genus $G$ and area $A$ is given by 
\bea 
Z_{ G, A  } = \sum_{ R }  ( Dim R )^{2-2G} e^{- g_{YM}^2 A C_2 ( R ) } 
\eea 
Of special interest to us is the zero area limit 
\bea 
Z_{ G } = \sum_{ R }  ( Dim R )^{2-2G} 
\eea 
$R$ runs over irreducible  representations of $SU(N)$ and $Dim(R)$
is the dimension of the representation.  
The results for manifolds with boundary are also known. 
Our main interest will be in the situation 
where $ \chi = 2-2G - B  \le  - 1 $.  The string theory interpretation 
of the ${ 1 \over N }$ expansion was developed in \cite{gt}.  
For earlier work on stringy aspects of 2DYM, see \cite{kazakov}. 
 For a review of the 
exact partition function, its ${1\over N}$ expansion, 
and the string theory interpretation see \cite{cmr}. 
$\Sigma ( G , B ) $ in this paper denotes a Riemann surface of 
genus $G$ with $B$ boundaries. 

The large $N$ expansion of these partition functions 
is described in terms of a coupling of a chiral partition 
function $Z^+$ with an anti-chiral partition function 
$Z^-$ \cite{gt}.  The chiral parition function is obtained by replacing
$ \sum_R \rightarrow \sum_n \sum_{ R  \vdash n }$
\bea\label{chiralexp}  
Z_{G }^+ = \sum_{ n =0}^{ \infty} \sum_{ R \vdash n }  ( Dim R )^{2-2G}     
\eea 
$R$ now runs over Young diagrams with $n$ boxes. 
Using Schur-Weyl duality, which relates the actions of $U(N)$ (or $SU(N)$)   
and $S_n$ in $V^{\otimes n }$,   this can be manipulated to give 
\bea\label{chirexp}  
Z_{G}^+ = \sum_{ n } { N^{n(2-2G)} \over n! }
 \sum_{ s_1, t_1 \cdots s_G , t_G \in S_n }
\delta_{n} ( \Omega_{n}^{2-2G } \prod_{i=1}^G s_i t_i s_i^{-1} t_i^{-1}  ) 
\eea 
where the ``chiral $\Omega$ factor''  $ \Omega_{n } $ is an element in the group algebra $ \C ( S_n ) $  
\bea 
\Omega_{n } = \sum_{ \sigma \in S_n  } N^{ C_{\sigma} - n } ~ \sigma 
\eea 
The $\delta_{n} $ is defined over $S_n$ by 
\bea\label{defdel}  
\delta_n ( \sigma ) = 1 \hbox { if }  \sigma = 1  \hbox{ and }  0 \hbox{ otherwise }  
\eea 
and extended over the group algebra $ \C ( S_n ) $  by linearity. 
 The expansion (\ref{chirexp})
 can be used to show that each order in the 
 $( { 1\over N }  )^{2g-2} $ expansion of $Z^+$ is a sum over equivalence classes 
of branched covers from a worldsheet Riemann surface of genus $g$ 
to the target $\Sigma_{ G}  $, so that we have a topological 
string theory with $g_s = { 1\over N }  $.   It is useful to define 
$ b(\sigma ) =  n -  C_{\sigma} $ which is the branching number
 of the permutation. The 
Riemann-Hurwitz formula 
\bea 
2g-2 = n ( 2G-2) + \sum_{ i } b( \sigma_i  )  
\eea 
gives the Euler character of the worldsheet for a 
branched cover of $ \Sigma_G$ with branchings $i$ described 
by $ \sigma_i$. So we see that  the power $N^{-b(\s ) } $ 
appearing in the $\Omega $ factor  is compatible with the 
interpretation of $ \Omega_{n}$ in terms of 
branch points with $g_s = {1\over N} $.
The $ \Omega_n $ can be written as $ 1 + \sum^{\prime} \sigma N^{-b(\s ) } $ 
and we have the expansion 
\bea\label{expOmeg}  
\Omega_{n}^{2-2G } = \sum_{ L=0}^{\infty } d(2-2G, L ) 
 \sum^{ \prime}_{\s_1 , .. \s_L} 
    \s_1 \s_2 \cdots \s_L ~ N^{- b ( \s_1 ) - b ( \s_2 ) \cdots ~ - ~ b( \s_L) }  
\eea 
where $ d(2-2G,L)$ is a binomial coefficient. 
The $L=0$ term is defined as $1$. 
 This factor $ d(2-2G,L)$,  related to the exponent $2-2G$  in $ \Omega_n^{2-2G} $, 
 is the  Euler character of the  configuration space of $L$ indistiguishable  points on $  \Sigma_G $. 
Along with the structure of Hurwitz spaces as  discrete 
fibrations over these configuration spaces, (\ref{chirexp}) and (\ref{expOmeg})
 are  used to   prove  that the $ Z_{G}^+ $ is a generating function 
of (orbifold) Euler characters of Hurwitz spaces of holomorphic maps from $\Sigma_g $ to 
$ \Sigma_{G } $. This is explained in section 5 of \cite{cmri} 
and reviewed in \cite{cmr}.

The complete expansion of the $Z_G$ takes a similar form 
\bea\label{nchirexp}  
Z_{G} = \sum_{ m,n =0 }^{\infty}  { N^{(m+n) (2-2G) } \over m! n! }
 \sum_{R \vdash m , S \vdash n  } ~~ 
 \sum_{ s_1, t_1 \cdots s_G , t_G \in S_m \times S_n }
\delta_{m,n}  ( \Omega_{m,n}^{2-2G } \prod_{i=1}^G s_i t_i s_i^{-1} t_i^{-1}  ) 
\eea 
The $ \delta_{m,n} $ is the delta function defined as in (\ref{defdel}) 
but for $ \C ( S_m \times S_n ) $. The ``coupled Omega factor''   $ \Omega_{m,n} $  is much more intricate 
than the ``chiral Omega factor'' $ \Omega_{n} $ but 
\bea 
\Omega_{ m ,n } = \Omega_m \Omega_n 
\left( 1 + \cO \left( { 1 \over N  } \right) \right)
\eea 
The detailed formula  and its interpretation in terms of  wordsheet geometry is reviewed in section 4.   
The complete expansion is interpreted in terms of 
maps from worldsheets which  have double points 
connecting two components which have branchings described 
by permutations in $S_m$ and $S_n$ respectively. 
The $m$ sheets map holomorphically to the target 
and the $n$ sheets map anti-holomorphically \cite{gt}.   More precisely 
$ Z_{G}$ generates  Euler characters of the appropriate moduli space of
maps (see section 10 of \cite{cmri}).

Recent work \cite{kimram} has used, in the context of branes and
 anti-branes in the 
AdS dual of 4D super-Yang Mills,  the Schur Weyl-duality between 
$U(N)$ acting on 
$ V^{ \otimes m } \otimes \bar V^{\otimes n } $ 
and the Brauer algebra $B(m,n)$. This algebra contains $\C(S_m \times S_n) $ 
as a sub-algebra, but also has additional generators corresponding to 
contractions between $V$ and $ \bar V$.  Another important property 
of $B(m,n)$ is that there is a map
 $ \Sigma  : B (m,n) \rightarrow \C(S_{m+n})$. 
This map is not a homomorphism but maps the natural  bilinear 
symmetric form on  $B(m,n)$ to a bilinear symmetric form on the 
group algebra $\C ( S_{m+n} ) $,   which can be calculated in terms of 
$\Omega_{m+n}$. The inversion of the form,  which is useful in constructing 
projection operators in the Brauer algebra \cite{ramthesis}, is conveniently 
done using $ \Omega_{m+n}^{-1} $. 
This can be used to derive a  formula 
for $ Dim R\bar S $  in terms of $S_{m+n}$ data \cite{kimram}. 
In this paper we will develop this further to 
 derive  a simple relation between $ \Omega_{m,n }^{-1}  $ and
 $ \Omega_{m+n}^{-1} $.
We then describe the implications for the string interpretation of 
the ${1\over N} $ expansion of 2DYM.

Section 2 derives the relation between $ \Omega_{m,n }^{-1}  $ and 
$ \Omega_{m+n}^{-1}$. In section 3 we use it to rewrite 
the complete ${1\over N} $  expansion of 2DYM for $ \Sigma ( G=2)$. 
In section 4 we give the geometrical interpretation.  
In section 5 we show that the same discussion carries over for 
 $\Sigma ( G) $.  The complete expansion (\ref{nchirexp}) 
can be interpreted in terms of holomorphic maps. As discussed in section 4,
for the case $ \Sigma ( G=1 , B=1 ) $ this is a straightforward 
consequence of the new formula for $ \Omega_{m,n}^{-1} $.  
In general it  requires a choice of cutting of $ \Sigma ( G ) $ 
into components of Euler character $-1$ i.e $3$-holed spheres 
or $1$-holed tori.

\section{New dimension formula and $\Omega $ factors }
In \cite{kimram}, we obtained a new formula 
for the coupled dimension, 
\begin{equation}
\frac{1}{Dim R\bar{S}}
=\frac{1}{d_{R}^{2}d_{S}^{2}}
\left(\frac{m!n!}{(m+n)!}\right)^{2}
\sum_{T}\frac{d_{T}^{2}}{DimT}g(R,S;T)
\label{newformula}
\end{equation}

We rewrite it using the formula for the 
Littlewood-Richardson (LR) coefficient 
\begin{equation}
g(R,S;T)=\frac{1}{d_{R}d_{S}DimT}tr_{m+n}((p_{R}\circ p_{S})p_{T})
\end{equation}
$p_R$ is a projection operator in $\C (S_m) $ (see Appendix A), 
$p_S $ is in $\C ( S_n ) $ and $ p_T $ in $\C ( S_{m+n}   ) $.  
We also use 
\begin{equation}
tr_{m+n}(\sigma)=N^{m+n}\delta_{m+n}(\Omega_{m+n}\sigma)
\end{equation}
and 
\begin{equation}
\frac{1}{(DimT)^{2}}
=\left(\frac{(m+n)!}{N^{m+n}d_{T}}\right)^{2}
\frac{\chi_{T}(\Omega_{m+n}^{-2})}{d_{T}}
\end{equation}
Then we have 
\begin{eqnarray}
\frac{d_{T}^{2}}{DimT}g(R,S;T)
&=&
\frac{d_{T}^{2}}{d_{R}d_{S}}\frac{1}{(DimT)^{2}}
tr_{m+n}((p_{R}\circ p_{S})p_{T}) \cr
&=&
\frac{1}{d_{R}d_{S}}
\left(\frac{(m+n)!}{N^{m+n}}\right)^{2}
\frac{\chi_{T}(\Omega_{m+n}^{-2})}{d_{T}}
tr_{m+n}((p_{R}\circ p_{S})p_{T}) \cr
&=&
\frac{1}{d_{R}d_{S}}
\left(\frac{(m+n)!}{N^{m+n}}\right)^{2}
tr_{m+n}(\Omega_{m+n}^{-2}(p_{R}\circ p_{S})p_{T}) \cr
&=&
\frac{1}{d_{R}d_{S}}
\left(\frac{(m+n)!}{N^{m+n}}\right)^{2}N^{m+n}
\delta_{m+n}(\Omega_{m+n}^{-1}(p_{R}\circ p_{S})p_{T}) 
\end{eqnarray}
%%%%%%%%%%%%
Therefore the formula (\ref{newformula}) can be brought to the form 
\begin{eqnarray}
\frac{1}{Dim R\bar{S}}
&=&\sum_{T}
\frac{1}{d_{R}^{3}d_{S}^{3}}
\frac{m!^{2}n!^{2}}{N^{m+n}}
\delta_{m+n}(\Omega_{m+n}^{-1}(p_{R}\circ p_{S})p_{T}) \cr
&=&
\frac{1}{d_{R}^{3}d_{S}^{3}}
\frac{m!^{2}n!^{2}}{N^{m+n}}
\delta_{m+n}(\Omega_{m+n}^{-1}(p_{R}\circ p_{S})) 
\label{1/dimRS}
\end{eqnarray}
where we have used $\sum_{T}p_{T}=1$. 
This can be written as 
\bea 
\frac{1}{Dim R\bar{S}} &=& \sum_{ \sigma \in S_m }   \sum_{ \tau  \in S_n } 
\frac { m! n! }{d_R^2 d_S^2 N^{m+n} }
 \chi_{R \otimes S } ( \sigma^{-1}  \otimes \tau^{-1}  )  
 \delta_{m+n}(  ( \sigma \otimes \tau ) \Omega_{m+n}^{-1} ) \cr 
&=&  \frac { m! n! }{d_R^2 d_S^2 N^{m+n} } \chi_{R \otimes S } 
( \Omega^{-1}_{m+n}  \vert_{S_m \times S_n } ) 
\eea 
$ \Omega^{-1}_{m+n}  \vert_{S_m \times S_n } $ is calculated  by 
expanding $ \Omega^{-1}_{m+n} $ as an element of the group algebra 
of $S_{m+n} $, and then restricting to the subgroup $ S_m \times S_n$. 
Comparing with the Gross-Taylor formula  in terms of the
 coupled-Omega factor we find 
that 
\bea\label{newcoupform}  \fbox{
$\displaystyle{
\Omega_{m,n}^{-1}  = \Omega_{m+n} ^{-1} \vert_{S_m \times S_n }
}$
}
\eea
As a simple example, we have 
\bea  
  \Omega_{2}^{-1} &=& \left( 1 - { 1 \over N^2 } \right)^{-1} 
  -  { \sigma \over N } \left( 1 - { 1 \over N^2 } \right)^{-1}  \cr 
  \Omega_{2}^{-1}\vert_{S_1 \times S_1} &=& 
  \left( 1 - { 1 \over N^2 } \right)^{-1} \cr 
  &=& \Omega_{1,1}^{-1} 
\eea 
In Appendix 
\ref{sec:omegachiralandcoupled}, other examples involving 
$m+n= 3,4 $ are illustrated.  An important point is that 
the relation   (\ref{newcoupform}) exists in the above 
simple form for $ \Omega_{m,n}^{-1} $ and not $ \Omega_{m,n}$. 
We cannot write $ \Omega_{m,n}$ as a projection of $ \Omega_{m+n}$. 
We can get  $ \Omega_{m,n}$ from $S_{m+n}$ data by using 
(\ref{newcoupform}) and then inverting after the projection. 
This is related to the fact that the new geometrical interpretation 
of the complete ${ 1 \over N } $ expansion which we propose, 
works best for $ \chi_{G,B}  = 2-2G - B \le -1 $.

\section{Genus 2 target }

\subsection{ genus $2$  : Partition function in terms of $S_{m+n }$   } 

We will prove that the complete $1 /N $ expansion of  $Z_{G=2}$ 
is given by 
\begin{eqnarray}\label{gen2first} 
Z_{G=2}
&=&\sum_{m,n} \sum_{ \alpha_1 \in S_m \times S_n  }
    { N^{-m-n}\over m! n! } \delta_{m+n}  ( \Omega_{m+n}^{-1} \Pi_1 \alpha_1  )  \cr 
&&    \sum_{ \alpha_2  \in S_m \times S_n  }     { N^{-m-n} \over m! n! } 
\delta_{m+n}  ( \Omega_{m+n}^{-1} \Pi_1 \alpha_2^{-1}  )  \cr 
&& \sum_{ \gamma \in   S_m \times S_n     } \delta_{m+n}  ( \alpha_1^{-1} \gamma
 \alpha_2 \gamma^{-1} ) 
 \eea 
Here $ \Pi_1 = \sum_{s,t \in S_m \times S_n } sts^{-1}t^{-1} $. 
The above formula corresponds to gluing two 1-holed tori to get
 a genus 2-surface (see figure \ref{fig:genus2onecut}). Boundary 
partition functions with symmetric group data and their gluing is 
reviewed in Appendix D.  

\begin{figure}
\begin{center}
 \resizebox{!}{4cm}{\includegraphics{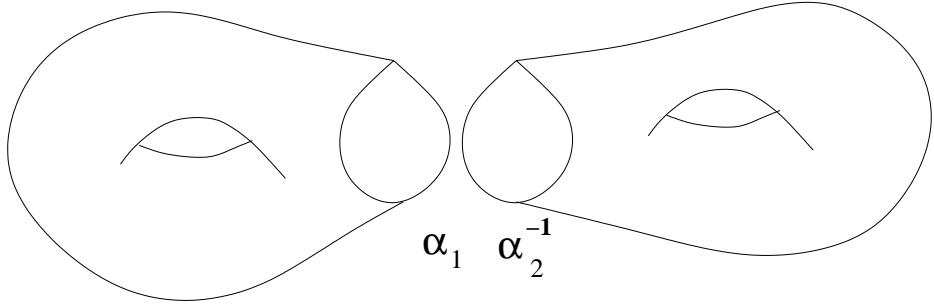}}
\caption{genus two from gluing two copies of $ \Sigma( G=1, B=1)$   }
 \label{fig:genus2onecut}
\end{center}
\end{figure}

We will also show 
\bea\label{gen2pants}  
Z_{G=2} &=&  
\sum_{m,n} \sum_{\alpha_1 ,  \alpha_2 ,  \alpha_3 \in S_m \times S_n  }
  { N^{-m-n} \over m! n! } \delta_{m+n} 
 ( \Omega_{m+n}^{-1} \alpha_1 \alpha_2 \alpha_3 ) \cr 
&& \sum_{\beta_1 ,  \beta_2 ,  \beta_3 \in S_m \times S_n  } 
 { N^{-m-n} \over m! n! } \delta_{m+n}  ( \Omega_{m+n}^{-1} \beta_1 \beta_2 \beta_3 ) \cr 
&& \sum_{\gamma_1 , \gamma_2 , \gamma_3  \in S_m \times S_n } 
       \prod_{i=1}^{3} \delta_{m+ n} ( \alpha_i^{-1} \gamma_i^{-1} \beta_i^{-1} \gamma_i ) 
\eea 
This corresponds to the gluing of two 3-holed
 spheres to get the genus $2$ curve as in Figure \ref{fig:genus2-3cuts}.

\subsection{ Derivations } 
Let us derive (\ref{gen2first}). We recall from \cite{gt} that 
\bea\label{gtexp} 
 Z_{G=2} = \sum_{m,n}   { N^{-2 (m+n) }  \over m! n! } \delta_{m,n}  ( \Omega_{m,n}^{-2}~  \Pi_1^2 )
\eea
We will 
use the abbreviation $ S_{m,n} \equiv  S_m \times S_n$. The delta function is 
over the group algebra of $S_{m,n}$. 
We can write this as 
\bea 
Z_{G=2} &=& \sum_{m,n}  { N^{- 2(m+n) }  \over m! n! } \delta_{m,n}  ( \Omega_{m,n}^{-1}~  \Pi_1 
~ \Omega_{m,n}^{-1}~  \Pi_1 ) \cr 
& =&  \sum_{m,n} ~~ \sum_{\alpha  \in S_{m,n}  }{ N^{- 2(m+n) }  \over m! n! }    \delta_{m,n}  ( \Omega_{m,n}^{-1} ~ \Pi_1~  \alpha ) 
~~ \delta_{m,n}    (   \Omega_{m,n}^{-1} ~ \Pi_1  ~ \alpha^{-1} ) \cr 
&=&  \sum_{m,n} ~~ \sum_{\alpha_1, \s_2   \in S_{m,n}  } { N^{- 2(m+n) }  \over m! n! }  
~~\delta_{m,n}  ( \Omega_{m,n}^{-1}~  \Pi_1  ~ \alpha_1 )  ~~\delta_{m,n}    (  \Omega_{m,n}^{-1}~ \Pi_1 ~ \alpha_2^{-1} )  
~~\delta_{m,n}  ( \alpha_1^{-1}  \alpha_2 ) \cr 
& = &   \sum_{m,n}~~  \sum_{\alpha_1, \s_2, \g    \in S_{m,n}  } { N^{- 2(m+n) }  \over  (m! n!)^2  }  
\delta_{m,n}  ( \Omega_{m,n}^{-1} ~\Pi_1 ~\g \alpha_1 \g^{-1}  ) ~~ \delta_{m,n}    ( \Omega_{m,n}^{-1}~ \Pi_1  ~\alpha_2^{-1} )  
~~\delta_{m,n}  ( \alpha_1^{-1}  \alpha_2 ) \cr 
& = &   \sum_{m,n}  ~~\sum_{\alpha_1, \s_2, \g    \in S_{m,n}  } { N^{- 2(m+n) }  \over  (m! n!)^2  }  
\delta_{m,n}  ( \Omega_{m,n}^{-1}~ \Pi_1 ~ \alpha_1   )  ~~\delta_{m,n}    (\Omega_{m,n}^{-1} ~\Pi_1 ~ \alpha_2^{-1}  )  
~~ \delta_{m,n}  (\gamma   \alpha_1^{-1}  \g^{-1}   \alpha_2 ) \cr 
& = &  \sum_{m,n}~~ \sum_{\alpha_1  \in S_{m,n}  } { N^{- (m+n) }  \over  m! n! } \delta_{m,n}  ( \Omega_{m,n}^{-1}~ \Pi_1  ~ \alpha_1   )  \cr 
   && \qquad \quad \sum_{\alpha_2  \in S_{m,n}  }{ N^{- (m+n) }  \over  m! n!}  \delta_{m,n}    (  \Omega_{m,n}^{-1} ~ \Pi_1 ~ \alpha_2^{-1} ) \cr 
  &&  \qquad \qquad  \sum_{ \g\in S_{m,n}}  ~~ \delta_{m,n}  (\gamma   \alpha_1^{-1}  \g^{-1}   \alpha_2 ) 
\eea 
The steps are each trivial. To get  to the fourth equality we have used the fact that 
the $ \Omega_{m,n}^{-1} ~ \Pi_1 $ is central in $S_{m,n}$. 
Now we will use  a simple rewriting of
$ \delta _{ m,n} ( \Omega_{m,n}^{-1}  B ) $   where 
$  B   \in  \C    ( S_m  \times S_n )  $. 
We know (\ref{newcoupform})  that $ \Omega_{m,n}^{-1} $ can be written
 as a projection of $ \Omega_{m+n}^{-1}   \in  \C    ( S_{m+n}  ) $. 
But when  we have $ \delta _{ m,n} ( \Omega_{m,n}^{-1 }   B ) $ with 
 $ B  \in  \C    ( S_m  \times S_n )  $, then the delta function 
can be rewritten as $ \delta _{ m,n} ( \Omega_{m,n}^{-1}  B ) 
= \delta_{m+n}  ( \Omega_{m+n}^{-1}  B ) $. 
The projection  is being performed by 
the $ \delta_{m+n}  $ on the group algbera  $ \C    ( S_{m+n}  ) $ and
 the fact that $B$ belongs to the subalgebra. 
Using this observation  we have 
\bea 
Z_{G=2} &&= \sum_{m,n}~~ \sum_{\alpha_1  \in S_{m,n}  } { N^{- (m+n) }  \over  m! n! } \delta_{ m+n }  ( \Omega_{m+n}^{-1}~ \Pi_1  ~ \alpha_1   )  \cr 
   && \qquad \quad \sum_{\alpha_2  \in S_{m,n}  }{ N^{- (m+n) }  \over  m! n!}  \delta_{m+n}    (  \Omega_{m+n}^{-1} ~ \Pi_1 ~ \alpha_2^{-1} ) \cr 
  &&  \qquad \qquad  \sum_{ \g \in S_{m,n} }  ~~ \delta_{m+n}  (\gamma   \alpha_1^{-1}  \g^{-1}   \alpha_2 ) 
\eea 
An important point is 
 that the replacement $  \delta_{ m,n} \rightarrow \delta_{m+n} ; 
\Omega_{m,n}^{-2}  \rightarrow  \Omega_{m+n}^{-2} $ 
cannot be done directly in (\ref{gtexp})  because $ \Omega_{m+n}^{-2}$ 
involves multiplying $  \Omega_{m+n}^{-1} \cdot  \Omega_{m+n}^{-1} $ 
with both being viewed as elements in $  \C    ( S_{m+n}  ) $, 
whereas $ \Omega_{m,n}^{-1} \cdot   \Omega_{m,n}^{-1} $ is a multiplication 
in $ \C    ( S_m  \times S_n )  $. 

We will now demonstrate (\ref{gen2pants}).  
Rewriting  (\ref{gtexp}) by expanding $ \Pi_1 $ 
\bea\label{gen2-3holeglue} 
 Z_{G=2} &=&\sum_{ s_1 , t_1 , s_2 , t_2 \in S_{m,n}  } 
{ N^{ -2m -2n} \over m! n! }  \delta_{m,n}  ( \Omega_{m,n}^{-2} ~ s_1 t_1 s_1^{-1} t_1^{-1}  ~ s_2 t_2 s_2^{-1} t_2^{-1}   )\cr 
& =& \sum_{ s_1 , t_1 , s_2 , t_2 \in S_{m,n}  } 
{ N^{ -2m -2n} \over m! n! }  \delta_{m,n}  ( \Omega_{m,n}^{-2}  ~ s_1 s_2  ~  t_1 s_1^{-1} t_1^{-1}  ~    t_2 s_2^{-1} t_2^{-1}   )\cr 
& =& \sum_{ s_1 , t_1 , s_2 , t_2 , s_3   \in S_{m,n}   } 
{ N^{ -2m -2n} \over m! n! }  \delta_{m,n}  ( \Omega_{m,n}^{-1}  ~  s_1 s_2  s_3  ) ~ \delta_{m,n}  ( \Omega_{m,n}^{-1}     ~  t_1 s_1^{-1}t_1^{-1}  ~    t_2 s_2^{-1} t_2^{-1} ~ s_3^{-1}  )\cr 
& =&  \sum_{ s_i , t_{i}, u_i   \in S_{m,n}  } 
{ N^{ -2m -2n} \over m! n! }  \delta_{m,n}  ( \Omega_{m+n}^{-1} ~  s_1 s_2  s_3  ) 
     ~ \delta_{m,n}  ( \Omega_{m+n}^{-1} ~      u_1  u_2       u_3  )\cr 
&& \qquad \qquad  \delta_{m,n}  ( t_1 s_1^{-1} t_1^{-1} u_1^{-1}   ) 
~ \delta_{m,n}  (  t_2 s_2^{-1} t_2^{-1} u_2^{-1}   ) 
~ \delta_{m,n}  ( u_3^{-1}  s_3^{-1}  ) \cr 
& =&  \sum_{   s_i , t_i , u_i   \in S_{m,n}  } 
{ N^{ -m -n} \over m! n! }  \delta_{m,n}  ( \Omega_{m+n}^{-1} s_1 s_2  s_3  )  ~~ 
  { N^{ -m -n} \over m! n! }     \delta_{m,n}  ( \Omega_{m+n}^{-1} u_1  u_2     u_{3}   )\cr 
&&  \qquad \qquad \delta_{m,n}  ( t_1 s_1^{-1} t_1^{-1} ~ u_1^{-1}   ) 
\delta_{m,n}  (  t_2 s_2^{-1} t_2^{-1}~  u_2^{-1}   ) 
\delta_{m,n}  ( t_3  s_3  t_3^{-1} ~ u_3^{-1}  )  
\eea 
To get to the last equality, we have inserted 
$1=\frac{1}{m!n!}\sum_{t_{3}\in S_{m,n}}t_{3}t_{3}^{-1}$ into the 
inside of the last $\delta_{m,n}$ and 
redefined  some variables. 
After a renaming $ s_i \rightarrow \alpha_i , u_i \rightarrow \beta_i ,
t_i \rightarrow \gamma_i $, this proves (\ref{gen2pants}).

\begin{figure}
\begin{center}
 \resizebox{!}{4cm}
{\includegraphics{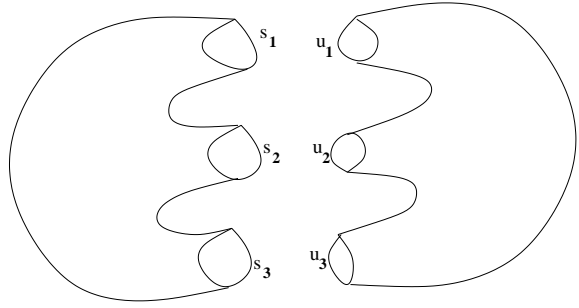}}
\caption{genus two from gluing two copies of $ \Sigma( G=0, B=3)$   }
 \label{fig:genus2-3cuts}
\end{center}
\end{figure}

\subsection{ Chiral form of complete ${1\over N} $ expansion for $ G=2 $  } 
The  chiral expansion of the partition function 
for genus $2$ can be written 
in the same form as either  (\ref{gen2first})  or (\ref{gen2pants}). 
We use the label $M$ for degree, and write for the chiral theory
the form corresponding to  the gluing of Figure \ref{fig:genus2-3cuts}. 

 \bea\label{gen2pantschir}  
Z^{+}_{G=2} &=& \sum_{M} 
 \sum_{\alpha_i  \in S_{M}   }  { N^{-M}  \over M! } \delta_M  ( \Omega_{M }^{-1} \alpha_1 \alpha_2 \alpha_3 ) \cr 
&& \sum_{\beta_i  \in S_{M}  }  { N^{-M}  \over M!  } \delta_M  ( \Omega_{M }^{-1} \beta_1^{-1}  \beta_2^{-1}  \beta_3^{-1}  ) \cr 
&& \sum_{\gamma_1 , \gamma_2 , \gamma_3  \in S_M  } 
       \prod_{i=1}^{3} \delta_M  ( \alpha_i^{-1} \gamma_i^{-1} \beta_i \gamma_i ) 
\eea 
To emphasize the similarity between (\ref{gen2pants}) and 
(\ref{gen2pantschir}) 
we can rewrite  (\ref{gen2pants}) as 
\bea\label{gen2chirobs} 
Z_{G=2} &=& \sum_{M} 
\sum_{\alpha_i  \in S_{M}   }  { N^{-M}  \over M! } \delta_M  ( \Omega_{M }^{-1} \alpha_1 \alpha_2 \alpha_3 ) \cr 
&& \sum_{\beta_i  \in S_{M}  }  {  N^{-M} \over M!   } \delta_M  ( \Omega_{M }^{-1} \beta_1^{-1}  \beta_2^{-1}  \beta_3^{-1}  ) \cr 
&& \sum_{\gamma_1 , \gamma_2 , \gamma_3  \in S_M  } 
       \prod_{i=1}^{3} \delta_M  ( \alpha_i^{-1} \gamma_i^{-1} \beta_i \gamma_i ) \cr 
&& \sum_{ H } \left({M! \over |H| }\right)^{2}  \prod_{i=1}^{3}   \delta_M  ( \alpha_i  { \bf 1 }_{H}   )  \delta_M  ( \beta_i { \bf 1 }_{H}   ) \delta_M  ( \gamma_i  { \bf 1 }_{H}  ) 
\eea 
where there is an additional sum over sub-groups $H = S_m \times S_n  $  in $ S_M \equiv S_{m+n} $ , 
with $M \ge  m , n \ge 0$. We have also defined 
$   { \bf 1 }_H  \equiv  \sum_{ \alpha \in H } \alpha $,  
 the projector onto the symmetric irrep of $H$. 
The delta functions  in the last line ensures that 
the permutations $ \alpha_i , \beta_i , \gamma_i $ are in the 
subgroup $H$.

\section{  Two holomorphic descriptions of the non-chiral expansion }

\subsection{ The Gross-Taylor coupled expansion: Worldsheets, nodes 
             and collision of branch points  } 

The coupled Omega factor has an expansion \cite{gt}   
\begin{eqnarray}\label{coupomegform} 
&&
\Omega_{m,n}=\sum_{\sigma^+  \in S_{m} } \sum_{\s^{-} \in S_{n}}
(\sigma^{+} \otimes \s^{-}  )P_{\s^+ \s^- } N^{C_{\s^+ }+C_{ \s^- }-(m+n)}
\cr
&&
P_{\s^+ \s^- }
=\prod _{j=1}\sum_{m}^{min(k_{j},l_{j})}
P_{k_{j},l_{j}}(m)\frac{1}{N^{2m}} \cr
&&
P_{k_{j},l_{j}}(m)=
\begin{pmatrix}
k_{j} \\
m
\end{pmatrix}
\begin{pmatrix}
l_{j} \\
m
\end{pmatrix}
m!(-j)^{m}
\end{eqnarray}
In the above,
 $j$ runs over the cycle lengths of $ \s^+, \s^- $. 
$k_j$ is the number of cycles of length $j $ in $\s^+ $ ; $l_j $ 
is the number of cycles of length $j$ in $\s^-$.

The coupled $\Omega_{m,n }  $ factor contains a sum of permutations 
weighted by polynomials in ${1\over N}  $. In  the chiral  $ \Omega_M$ 
factor, each permutation $ \sigma $ is weighted just by $N^{-b(\sigma) } $ 
where $ b( \sigma ) $ is the branching number of the permutation. 
In the case of the coupled $\Omega_{m,n} $ factor 
$ \sigma \in S_{m } \times S_n  $, written as $ \sigma^{+} \otimes \sigma^- $ 
to emphasize the product form, has a leading coefficient which 
is just the sum of branching numbers of the $  \sigma^{+} $ and  
$  \sigma^{-} $. The subleading terms have an elegant  combinatoric 
interpretation discovered  in \cite{gt}. They  
can be interpreted in terms of double points on the worldsheet, 
joining ramification points.  The rule is that the double points 
can only connect ramification points of the same order, which can be zero.  
Each such double point is accompanied by a weight of ${ (-j)  \over N^2 }  $ 
where $j$ is the order of the ramification. The factor of ${1 \over N^2 } $  
accounts precisely for the change in Euler character of the worldsheet 
upon introduction of such a double point.  A local model 
of  the double point and the map was given in \cite{cmri}. 
The  branching described by $ \sigma^{+} \in S_m  $ is taken to 
be that of a holomorphic map, and the branching described 
by $ \sigma^{-}$ is anti-holomorphic. 

It is instructive to consider an orientation reversal on $ \sigma^{-} $ 
so that both $ \sigma^{+} , \sigma^-$ describe holomorphic maps. 
Then we can ask how the double points of the coupled expansion 
arise from the collision of branch points. For example consider a 
double cover over the sphere which is branched with two branch points 
each corresponding to the permutation $ (12) $. After collision 
the monodromy is just the identity permutation, with branching number zero. 
The Euler character of the worldsheet has not changed. This is consistent 
with a double point arising from the collision, which joins 
two points of trivial ramification. If we take a collision 
of branch points described by $ (12 \cdots  j ~ j+1 ) $ and 
$ (j ~ j+1 \cdots 2j ) $, the resulting permutation is 
$ (1 \cdots  j ) ( j+1 \cdots  2j)$  with branching number two less than 
the sum of branching numbers of the collising permutations, so a 
double point has been created. So all the double points 
of the type arising in the coupled expansion can occur 
from the collision of branch points, i.e at the boundaries 
of Hurwitz space.  The most general collision of branch 
points can produce more complicated 
singularities, which do not all occur in the coupled expansion. 
For further comments on the collision of branch points 
and its relevance to the the identity (\ref{newcoupform})   
see the end of Appendix C.

By generalizing the argument of the chiral sector to the coupled 
case (section 10 of \cite{cmri}) 
it can be shown that the complete ${1\over N} $ expansion of 2DYM 
computes Euler characters of holo-anti-holo maps,  or after the orientation 
reversal just holomorphic  maps, from worldsheets that can have double points 
according to the rules described above. This was called   the moduli space 
of ``degenerating coupled covers'' in \cite{cmri}.

\subsection{ New holomorphic  interpretation 
in the case   $ \Sigma ( G=1 , B=1)  $ }

The simplest case where we can see the new interpretation based 
on (\ref{newcoupform}) is for 
2DYM on $ \Sigma ( G=1 , B=1)  $. 
In this case we have boundary 
observables specified by choosing an integer $n$ 
and a conjugacy class $T$ of 
 $ S_m \times S_n $ (see Appendix D for a quick review and \cite{cmr,ramwil}
 for more details)  
\bea 
Z ( G=1 , B=1 ; T  ) 
= \sum_{ \alpha \in T }  \sum_{ s,t \in S_m \times S_n } 
{ N^{-m-n}  \over m! n! }  \delta_{m,n} ( \Omega_{m,n}^{-1} s t s^{-1} t^{-1} \alpha ) 
\eea 
This partition function can be re-written as 
\bea 
Z ( G=1 , B=1 ; T  ) 
= \sum_{ \alpha \in T }  \sum_{ s, t \in S_m \times S_n } 
{  N^{-m-n}  \over m! n! }  \delta_{m+n} ( \Omega_{m+ n}^{-1} 
s t s^{-1} t^{-1} \alpha ) 
\eea 

The first expression can be expanded and interpreted as 
an Euler character of the moduli space of ``degenerating 
coupled covers'' \cite{cmri}. 
The second expression can be expanded
\bea 
&& Z ( G=1 , B=1 ; T  ) 
= \sum_{ \alpha \in T }  \sum_{ s, t \in S_m \times S_n } 
{ N^{-m-n}  \over m! n! }  \sum_{L=0}^{\infty} 
 d ( -1 , L )  \cr 
&& \sum_{ \sigma_1 \cdots \s_L \in S_{m+n} }^{ \prime } N^{-b(\s_1) - b(\s_2 ) - \cdots - b ( \s_L ) }    \delta_{m+n} 
( \s_1 \cdots \s_L  
s t s^{-1} t^{-1} \alpha ) 
\eea 
Since the permutations $ \sigma_1 \cdots \sigma_L$ are in $S_{m+n}$ 
the corresponding branch points can permute any of the $m+n$ sheets of the cover among 
each other. The power of $N$ is consistent with these branching numbers. 
 So these are  holomorphic maps of degree $ m+n$.  
The terms with fixed $L$  can be interpreted in terms  
of a moduli space of  holomorphic maps to $ \Sigma(G=1,B=1)$ 
with $L$ branch points with boundary permutation 
in the conjugacy class $T$.  The cycle lengths  of $T$ correspond to winding
 numbers of  strings at the boundary. The binomial factor $ d(-1,L) =(-1)^L$ 
is the Euler character of the configuration space 
 of $L$ indistinguishable points on $  \Sigma(G=1,B=1)$. 
Hence the $Z ( G=1 , B=1 ; T  )$ is a generating 
function for the  Euler character of the space of holomorphic 
maps with fixed string winding numbers at the boundary.

\subsection{ Holomorphic maps,  sums over $H$-monodromies 
 along markings : No singular worldsheets  }

We now describe  the holomorphic interpretation for closed 
target spaces, for simplicity in the case of $ \Sigma ( G=2) $. 
It will be clear that  the same ideas generalise to  
closed Riemann surfaces of any genus $G$.

The chiral expansion is a 
sum over $M$, which corresponds to the degree of the map
from worldsheet $ \Sigma_g $ to the target $ \Sigma (G)$. 
The chiral partition function $Z^+_{G=2}$ can be derived 
by gluing partition functions on one-holed tori or 
3-holed spheres, e.g.  (\ref{gen2pantschir}).
The final expression is  independent of the choice of decomposition     
into $ \chi=-1$ components as is manifest in (\ref{chirexp}).
For each degree, the data at each boundary required to 
specify the boundary partition function is a conjugacy class 
in $S_M$. In the gluing procedure, we sum over all conjugacy classes
in $S_M$ and subsequently sum over $M$.

In the formulae developed above (\ref{gen2first}) (\ref{gen2pants}), 
for the complete ${ 1\over N }$ expansion, the gluing 
procedure is generalised. It involves summing over  subgroups $H = 
S_m \times S_n$ of $S_{M=m+n}$. For each choice of $H$, we consider 
boundary partition functions depending on a conjugacy class in 
$H$, and the $s,t$ monodromies in the subgroup as well. 
The boundary permutations and the $s,t$ monodromies are 
summed over $H$. The branch points coming from expanding 
$ \Omega_{m+n}^{-1} $ however are general permutations 
in $S_M$. The branch points can permute any of the $M$ 
sheets, so these are holomorphic maps of degree $M$. 
For example, from  (\ref{gen2first}) we have 
\bea 
Z_{G=2}
&=&\sum_{m,n} \sum_{ \alpha_1 \in S_m \times S_n  }
    { N^{-m-n}\over m! n! }  \sum_{L_1=0}^{\infty}  ~~~ 
\sum_{ \s_1 , \cdots , \s_{L_1} \in 
S_{m+n}  }^{\prime}   d ( -1 , L_1 )   ~~ 
 N^{-b(\s_1) -b(\s_2) \cdots - b (\s_{L_1}) }   
 \cr 
&&   \hskip.2in  \sum_{ \alpha_2  \in S_m \times S_n  }^{\prime} 
     { N^{-m-n} \over m! n! }  \sum_{L_2=0}^{\infty} ~~~~ 
\sum_{ \tau_1 , \tau_2 , \cdots \tau_{L_2} \in S_{m+n}   }^{\prime}  
  d ( -1 , L_2 ) ~~  N^{-b(\tau_1) -b(\tau_2) \cdots - b (\tau_{L_2}) }
  \cr 
&&\hskip2in \sum_{ s, t \in S_m \times S_n} 
 \delta_{m+n}  ( \s_1 \s_2 \cdots \s_{L_1}   s t s^{-1} t^{-1} 
 \alpha_1  ) \cr 
&& \hskip2in  \sum_{ \tilde s, \tilde t \in S_m \times S_n} 
\delta_{m+n}( \tau_1 \tau_2 \cdots \tau_{L_2}   { \tilde s}  {\tilde t }
 {\tilde s}^{-1} {\tilde t}^{-1} 
                     \alpha_2^{-1}  ) \cr 
&& \hskip2.3in \sum_{ \gamma \in   S_m \times S_n  } \delta_{m+ n}  ( \alpha_1^{-1} \gamma
 \alpha_2 \gamma^{-1} ) 
\eea 
  The binomial coefficients  $d ( -1 , L_1 ) =  (-1)^{L_1}  $ and 
$ d ( -1 , L_2 ) =  (-1)^{L_2}$ in the expansion of
 the  $ \Omega_{m+n}^{-1} $ factors  are Euler characters  
 the configuration spaces of points on each $\chi=-1$ component
(see \cite{cmri} for a quick review of these Euler characters).  
Hence we can interpret in terms of the Euler character of 
a moduli space of holomorphic maps where the branch points can move 
over these components.  There are no singular wordsheets in this 
interpretation. All the branch points coming from expanding the 
 $ \Omega$ factors are weighted with powers of $N$ according to their 
 branching number.  

The complete partition function can be written in the suggestive form of 
an insertion, involving an additional  sum over subgroups $H$
of $S_M$, in the chiral partition function  (\ref{gen2chirobs}). 
In  the next section we will find the higher genus analogs of the 
 formulae in section 3.  The possible  implications of  
(\ref{gen2chirobs})  in terms of observables in a 
 topological string theory of holomorphic maps will be discussed 
 in the general genus case in section 5.

To summarize, the complete ${ 1 \over N } $ expansion of 
2DYM as given in (\ref{nchirexp}) can be interpreted as
a generating function of Euler characters of moduli spaces of 
holomorphic maps in two different ways. 
In one interpretation \cite{gt,cmr}, based on the formula 
for  $ \Omega_{m,n}^{-1}$ in \cite{gt},  there are  worldsheets  with double 
points and branch points which can  wander all over $ \Sigma ( G)$. 
The standard interpretation involves holomorphic and anti-holomorphic 
maps on different components joined at double points, but 
we can get a corresponding holomorphic moduli space by an orientation reversal 
on the anti-holomorphic component. 
A new interpretation using $ \Omega_{m+n}^{-1} $ 
follows directly from unravelling the 
consequences of (\ref{newcoupform}). In this interpretation there 
are no worldsheet double points and the branch points 
are free to move on the $ \chi=-1$  components of a decomposition 
of $ \Sigma ( G)$  fixed by choosing some 
markings on the Riemann surface. By generalising the way we lift 
the gluing together of spacetime $ \Sigma ( G ) $  along the markings 
to the gluing  of spaces  of maps from worldsheets to 
$ \Sigma ( G)$, we are able to get rid of the 
worldsheet double points.  Since the expressions 
(\ref{gen2first}) (\ref{gen2pants}) are derived from 
(\ref{nchirexp}), it is clear that the generalized gluing is 
compatible with a well-defined partition function on $ \Sigma ( G ) $, 
{\it  independent of the  choice of markings }  which separate $ \Sigma ( G ) $
into components of $\chi = -1 $. 
The equivalence of different descriptions which 
give rise to the same expansion in the string coupling $g_s = { 1\over N } $ 
is reminiscent of $T$-duality. A natural question is whether 
the equivalence of Euler characters of the two different moduli 
spaces described above is a duality that can be studied by
physical methods from the worldsheet point of view \footnote{We thank 
R. de Mello Koch for a discussion on this point. }.  

Given the crucial role played by the generalised gluing,  it 
would be interesting to investigate the generalisation 
 arising   in (\ref{gen2chirobs}),  involving a summation over 
subgroups $H$ of a group $\cal{ G} $, in the context of general 
two dimensional  topological field theories of a finite group $\cal{G}$. 
While ${\cal{G}} = S_M $ is of special interest in the large $N$ 
expansion of 2DYM due to  the Hurwitz connection between branched covers 
and $S_M$, other finite groups might be   of interest in connection with 
the topological sector of  CFTs with  orbifold target spaces. 

\section{ General Genus } 

\subsection{ $Z_G $ in terms of $ \delta_{m+n} $ } 

The following fomula is derived in Appendix B.1
\bea 
Z_{G}  
&=& \sum_{ m,n }  ~ \sum_{ \alpha_i \tilde \alpha_i , \beta_i , \tilde
 \beta_i  \in S_m \times S_n } 
\prod_{i=1}^{G}  {  N^{-m-n } \over m! n! }    \delta_{m+n}  ( 
  \Omega_{m+n}^{-1}   \Pi_1 \alpha_i^{-1}  ) \cr 
 && {  N^{-m-n } \over m! n! }  
 \delta_{m+n} ( \Omega_{m+n}^{-1} \tilde \alpha_1
 \tilde \alpha_2   \beta_1 ) 
{  N^{-m-n } \over m! n! }  
 \delta_{m+n} ( \Omega_{m+n}^{-1} \tilde \beta_1^{-1}  \tilde \alpha_3 \beta_2 ) \cr 
&& {  N^{-m-n } \over m! n! } 
\delta_{m+n} (  \Omega_{m+n}^{-1} \tilde \beta_2^{-1}  \tilde \alpha_4 \beta_3 )
 \cdots {  N^{-m-n } \over m! n! } \delta_{m+n} (    \Omega_{m+ n}^{-1}\tilde  
\beta_{G-3}^{-1}  \tilde \alpha_{G-1}   \tilde  \alpha_G )    \cr 
&& \sum_{ \gamma_i , \epsilon_i \in S_m \times S_n }  \prod_{i=1}^{G} \delta_{m+n } ( \alpha_{i} 
\gamma_i \tilde \alpha_i^{-1}  \g_i^{-1}   ) 
 ~~ \prod_{i=1}^{G-3}  \delta_{m+n  } ( \beta_{i}^{-1}  \e_i \tilde \beta_i \e_i^{-1}   )    
\eea 
The corresponding choice of markings that
 separate $ \Sigma  ( G ) $ into $\chi=-1$ components 
 is shown in  Figure \ref{fig:genG1ht}. 
 We have $G$ copies of 1-holed tori 
glued to 3-holed spheres by permutations $ \alpha_i$. There are $G-2$
 copies of 
the 3-holed spheres. 
\begin{figure}
\begin{center}
 \resizebox{!}{6.5cm}{\includegraphics{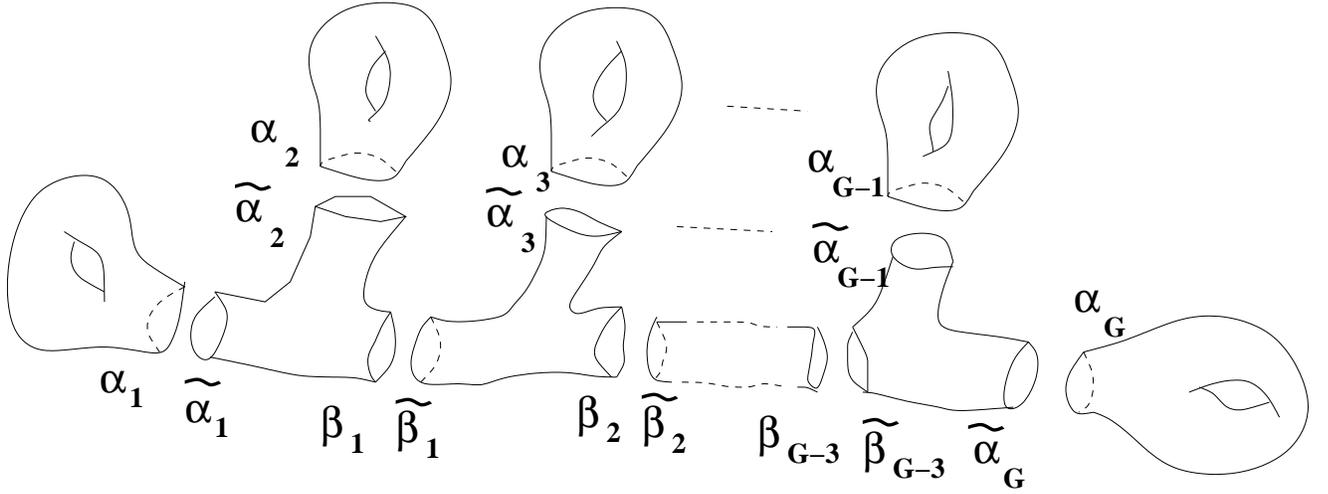}}
\caption{genus $G$ from gluing $G$ copies of $ \Sigma( G=1, B=1)$
 and $G-2$ copies of $ \Sigma( G=0,B=3) $    }
 \label{fig:genG1ht}
\end{center}
\end{figure}

Another way to write the partition function employs the 
cutting of genus $G$ into $2G-2$ copies of $ \Sigma ( G=0 , B=3) $
as in Figure \ref{fig:genGpants}.

\bea\label{genGpantseq}  
Z_{G} 
&&  = \sum_{ m , n } \sum_{s_i,t_i,u_i, v_i , w_i ,\tv_i ,\tw_i,\gamma_{i}, \e_i  \in S_m \times 
S_n }
 { N^{-(m+n) } \over m! n!  } \delta_{m+n} ( \Omega_{m+n}^{-1} s_1 s_2 v_1 )
 ~~ { N^{-(m+n) }      \over m! n!  } 
\delta_{m+n} ( \Omega_{m+n}^{-1} u_1 u_2 w_1 ) \cr 
 &&  \hskip1in { N^{-(m+n) } \over m! n!  } \delta_{m+n} 
 ( \Omega_{m+n}^{-1}  \tv_1^{-1}  s_3 v_2 ) ~~  { N^{-(m+n) }    
  \over m! n!  } \delta_{m+n} ( \Omega_{m+n}^{-1} \tw_1^{-1}  u_3 w_2 ) \cr 
&& \qquad  \hskip1.8in  \vdots \hskip1.5in \vdots \cr 
&&  \hskip1in  { N^{-(m+n) }      \over m! n!  } \delta_{m+n} ( \Omega_{m+n}^{-1} 
 \tv_{G-3}^{-1} s_{G-1} v_{G-2} )
   { N^{-(m+n) }      \over m! n!  }
\delta_{m+n} ( \Omega^{-1}  \tw_{G-3}^{-1} s_{G-1} w_{G-2} ) \cr 
&&  \hskip1in  { N^{-(m+n) }      \over m! n!  } \delta_{m+n} ( \Omega_{m+n}^{-1}
 \tv_{G-2}^{-1} s_G s_{G+1} )  { N^{-(m+n) }      \over m! n!  }  \delta_{m+n}
 ( \Omega_{m+n}^{-1} \tw_{G-2}^{-1} u_G u_{G+1} ) \cr 
&&  \hskip1in \prod_{i=1}^{G+1} \delta_{m+n} ( u_i t_i s_i^{-1} t_i^{-1} )  
  \prod_{ i=1}^{G-2}  \delta_{m+n} ( v_i^{-1}  \g_i  \tv_i   \g_i^{-1} ) 
                          \prod_{ i=1}^{G-2}  \delta_{m+n} ( w_i^{-1}  \e_i  \tw_i   \e_i^{-1} )         
\eea

We have introduced  the gluing permutations 
$ \gamma_i , \epsilon_i $ so that we  
get, for each component of Euler character $-1$ a
 boundary partition function with standard normalisation. 
We can therefore write 
\bea 
Z_{G}  &= &\sum_{m,n}   ~~~ \sum_{s_i,t_i,u_i, v_i , w_i ,\tv_i ,\tw_i,\gamma_{i},
\e_{i}
 \in S_{m}\times S_n } \cr
&& Z ( G=0, B=3 ; s_1 s_2 v_1 ) Z ( G=0, B=3 ;u_1 u_2 w_1 ) \cr 
&& Z ( G=0 , B=3 ;  \tv_1^{-1}  s_3 v_2 )  Z ( G=0, B=3  ; 
        \tw_1^{-1}  u_3 w_2 ) \cr 
&& \qquad  \hskip1in  \vdots \hskip1.5in \vdots \cr 
&& Z ( G=0 , B=3 ;  \tv_{G-3}^{-1} s_{G-1} v_{G-2} ) 
 Z ( G=0 , B=3 ;   \tw_{G-3}^{-1} s_{G-1} w_{G-2} ) \cr 
&&    Z ( G=0 , B=3 ;  \tv_{G-2}^{-1} s_G s_{G+1} )  
                  Z ( G=0 , B=3 ;  \tw_{G-2}^{-1} u_G u_{G+1} ) \cr 
&& \prod_{i=1}^{G+1} \delta_{m+n} ( u_i t_i s_i^{-1} t_i^{-1} )  
  \prod_{ i=1}^{G-2}  \delta_{m+n} ( v_i^{-1}  \g_i  \tv_i   \g_i^{-1} ) 
\prod_{ i=1}^{G-2}  \delta_{m+n} ( w_i^{-1}  \e_i  \tw_i   \e_i^{-1} )  \nn\\ 
\eea

\begin{figure}
\begin{center}
 \resizebox{!}{8cm}{\includegraphics{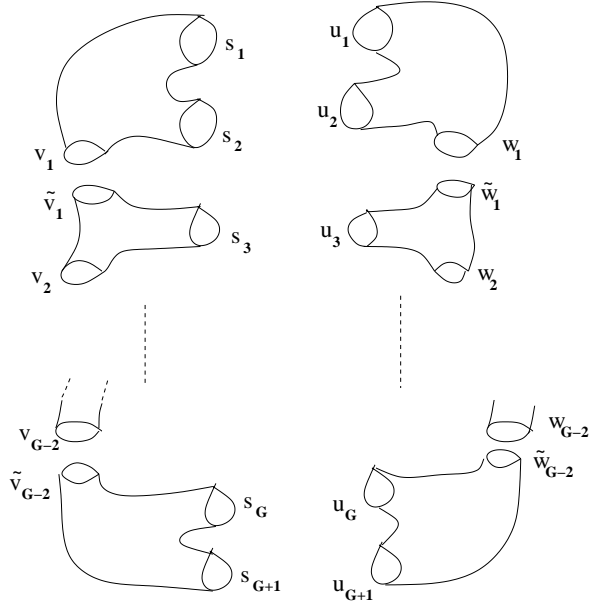}}
\caption{genus $G$ from gluing $2G-2$ copies of $ \Sigma( G=0, B=3)$   }
 \label{fig:genGpants}
\end{center}
\end{figure}

\subsection{ Chiral form of complete ${1\over N} $ expansion for general $G$. } 

The chiral partition function can also be written 
in a way to emphasize the construction of the genus $G$ surface by gluing 
pants diagrams
\bea\label{chiralG}  
Z_{G}^{+} &=&  \sum_M \sum_{s_i,t_i,u_i, v_i , w_i ,\tv_i ,\tw_i, \g_i, \e_i 
 \in S_M   } 
 Z^+ ( G=0, B=3 ; s_1 s_2 v_1 ) Z^+ ( G=0, B=3 ;u_1 u_2 w_1 ) \cr 
&&   Z^+ ( G=0 , B=3 ;  \tv_1^{-1}  s_3 v_2 )  Z^+ ( G=0, B=3  ; 
        \tw_1^{-1}  u_3 w_2 ) \cr 
&& \qquad  \hskip1in  \vdots \hskip1.5in \vdots \cr 
&& Z^+ ( G=0 , B=3 ;  \tv_{G-3}^{-1} s_{G-1} v_{G-2} ) 
     Z^+ ( G=0 , B=3 ;   \tw_{G-3}^{-1} s_{G-1} w_{G-2} ) \cr 
&& Z^+ ( G=0 , B=3 ;  \tv_{G-2}^{-1} s_G s_{G+1} )  
                  Z^+ ( G=0 , B=3 ;  \tw_{G-2}^{-1} u_G u_{G+1} )\cr 
&& \prod_{i=1}^{G+1} \delta_{M } ( u_i t_i s_i^{-1} t_i^{-1} )  
  \prod_{ i=1}^{G-2}  \delta_{M } ( v_i^{-1}  \g_i  \tv_i   \g_i^{-1} ) 
\prod_{ i=1}^{G-2}  \delta_{M } ( w_i^{-1}  \e_i  \tw_i   \e_i^{-1} )  \nn\\
\eea 

To emphasize the similarity between the full partition function 
and the chiral one, we write  
\bea\label{chiralGinsert} 
Z_{ G } 
 &=&  \sum_M \sum_{s_i,t_i,u_i, v_i , w_i ,\tv_i ,\tw_i, \g_i , \e_i \in S_M   } 
 Z^+ ( G=0, B=3 ; s_1 s_2 v_1 ) Z^+ ( G=0, B=3 ;u_1 u_2 w_1 ) \cr 
&&Z^+ ( G=0 , B=3 ;  \tv_1^{-1}  s_3 v_2 )  Z^+ ( G=0, B=3  ; 
        \tw_1^{-1}  u_3 w_2 ) \cr 
&& \qquad  \hskip2in  \vdots \hskip1.5in \vdots \cr 
&& Z^+ ( G=0 , B=3 ;  \tv_{G-3}^{-1} s_{G-1} v_{G-2} ) 
 Z^+ ( G=0 , B=3 ;   \tw_{G-3}^{-1} s_{G-1} w_{G-2} ) \cr 
&&     Z^+ ( G=0 , B=3 ;  \tv_{G-2}^{-1} s_G s_{G+1} )  
                  Z^+ ( G=0 , B=3 ;  \tw_{G-2}^{-1} u_G u_{G+1} )\cr 
&&  \prod_{i=1}^{G+1} \delta_{M} ( u_i t_i s_i^{-1} t_i^{-1} )  
  \prod_{ i=1}^{G-2}  \delta_{M} ( v_i^{-1}  \g_i  \tv_i   \g_i^{-1} ) 
\prod_{ i=1}^{G-2}  \delta_{M} ( w_i^{-1}  \e_i  \tw_i   \e_i^{-1} )  \cr 
&& \sum_{ H }   \left(\frac{M!}{|H|}\right)^{2G-2}
\prod_{i=1}^{G+1 } \delta_{M} ( s_i { \bf 1}_H  )
    \delta_{M} ( u_i { \bf 1}_H  )  \delta_{M} ( t_i { \bf 1}_H  ) 
    \cr
&&
\prod_{i=1}^{G-2 }
 \delta_{M} ( v_i { \bf 1}_H  ) \delta_{M} ( \tv_i { \bf 1}_H ) 
\delta_{M} ( w_i { \bf 1}_H  ) \delta_{M} ( \tw_i { \bf 1}_H  )
\delta_{M} ( \gamma_i { \bf 1}_H  )
\delta_{M} ( \e_i { \bf 1}_H  )
 \nn \\ 
\eea 

The sum over $H$ is the sum over $ S_m \times S_n$ subgroups, 
with $ 0 \le m, n  \le M $ and $ m+n = M $.    
We have defined $  { \bf 1}_H  \equiv  \sum_{ \sigma \in H } \sigma$. 
When $H$ is restricted to be $ S_M $, i.e $ (m,n) = (0,M)$ or $( M,0) $, 
we have the   standard chiral partitions functions. 
 The expression suggests an interpretation 
within a topological string 
theory of holomorphic maps with target $ \Sigma (  G ) $,
of the complete $ { 1\over N }$ expansion of 
 the 2DYM partition function,  
in terms of the insertion of an appropriate observable corresponding to 
\bea 
 &&\sum_{ H } \prod_{i=1}^{G+1 } \delta_{M} ( s_i { \bf 1}_H  )
    \delta_{M} ( u_i { \bf 1}_H  ) 
     \delta_{M} ( t_i { \bf 1}_H  ) \cr
     &&
   \prod_{i=1}^{G-2 }  
 \delta_{M} ( v_i { \bf 1}_H  ) \delta_{M} ( \tv_i { \bf 1}_H ) 
\delta_{M} ( w_i { \bf 1}_H  ) \delta_{M} ( \tw_i { \bf 1}_H  ) 
\delta_{M} ( \gamma_i { \bf 1}_H  )
\delta_{M} ( \e_i { \bf 1}_H  ) 
\eea 
Identifying such an obervable in terms of classes on the Hurwitz space 
of holomorphic maps, or in terms of the pull-back of these classes 
to the moduli space of worldsheet complex structures 
would be the next step in developing the holomorphic description 
of the full ${1\over N} $ expansion of 2DYM.
If the classes on $\overline { \cM}_{g,n}  $
can be  expressed in terms of the 
Mumford-Morita  classes with intersection numbers 
computed by 2D quantum gravity \cite{wit2D}, 
this could lead to new connections 
between 2DYM and integrable equations.  This would give a 
concrete way to explicitly  compute the 
terms in the ${1\over N} $ expansion directly from wordsheet 
topological string methods.

%%%%%%%%%%%%%%%%%%%%%%%%%%%%%%%%%%%%%%%%%%%%%%%%%%%%%%%%%%%%%%%%%%%%%%%%%%%%%
%%%%%%%%%%%%%%%%%%%%%%%%%%%%%%%%%%%%%%%%%%%%%%%%%%%%%%%%%%%%%%%%%%%%%%%%%%%%%

%%%%%%%%%%%%%%%%%%%%%%%%%%%%%%%%%%
\section{Summary  and Outlook } 

By developing results in \cite{kimram}  
we have found an expression  (\ref{newcoupform})
 for the coupled  inverse Omega factor $ \Omega_{m,n}^{-1} $ 
of \cite{gt} in terms of a projection of the chiral inverse 
Omega factor $\Omega_{m+n}^{-1} $.  The latter has a simple 
interpretation in terms of branch points. 
This has allowed us to write the complete ${1\over N} $ expansion 
of the partition function of 2DYM theory,  with $SU(N)$ gauge group,  
as an insertion of an observable  in the chiral partition function (see 
(\ref{gen2chirobs}) and (\ref{chiralGinsert})). 
The  chiral form of the complete ${1\over N}  $ expansion uses a 
choice of markings which separate the target space $ \Sigma(G)$ into 
components  of Euler character $-1$. i.e 3-holed spheres or 1-holed tori. 
The partition function  does not depend on the choice of markings. 
The difference between the chiral expansion and the complete one is 
simply  in the choice of gluing factors at the markings.
The complete expansion has an additional sum over subgroups
 $ S_m \times S_n $ of $S_M$ where $M=m+n$ 
is the degree of the map from worldsheet to 
target.  
The geometrical interpretation  of the coupled $\Omega_{m,n}^{-1}  $ factor
involves worldsheets with double points which can arise 
from collision of branch points. The expression in terms of the 
chiral $\Omega_{m+n}^{-1}  $ factor allows a geometrical interpretation 
with smooth worldsheets  $ \Sigma ( g ) $ without double points
mapping to the target space. In particular 
we have  an equality of the Euler character
of a space of holomorphic maps from worldsheets which can have nodes and the 
Euler character of a space of holomorphic maps from smooth worldsheets.

Several extensions of these results are worth investigating. 
Incorporating finite area $A$ or changing the gauge group 
from $SU(N)$ to $U(N)$ can be done trivially. The latter 
involves an extra sum over a $U(1)$ charge. The dimensions 
of irreps are unaffected by tensoring with the $U(1)$ representations, 
 so there is no non-trivial modification. In the bulk of this paper 
we have used $SU(N)$ rather than $U(N)$ because the 
main points about the chiral reformulation can be made 
at zero area and zero theta parameter in the former case. 
In the case of $U(N)$ we have to include the area or 
the theta parameter to control the infinite sum over 
$U(1)$ irreps. The generalisation of the large $N$ expansion of 2DYM for the 
gauge groups $ O(N)$ and $Sp(N)$  is known \cite{nrs,ram}. 
This expansion involves worldsheets with nodes, and the additional 
feature of possible non-orientability. Is there a rewriting 
of the $\Omega $ factors of $O(N), Sp(N)$ which allows us to 
map the partition function to one involving worldsheets that do not 
involve nodes ? In the $U(N)$ or $SU(N)$ theory, there are non-perturbative 
sectors which have an interpretation of terms of splitting fermi seas
\cite{dgov}. These theories also admit $q$-deformations which 
have a string interpretation in terms of strings with Calabi-Yau targets 
which are direct sums of line bundles over $\Sigma(G)$ \cite{aosv} 
(see also subsequent work \cite{ccgpsz,abms}). 
The chiral ${1\over N} $ (more precisely $1/[N]$ where $[N]$ is a 
$q$-number) expansion for these theories have been worked out
\cite{drt}. It would be interesting to work out the chiral formulation 
of the full expansion in the $q$-deformed and non-perturbative sectors. 
 An obvious  question is to find how to 
express the observable inserted in (\ref{gen2chirobs}) 
(\ref{chiralGinsert}) 
in terms of the balanced topological string proposed 
 as the worldsheet string for chiral 2DYM    (\cite{cmri,dijmoo}).

The result (\ref{newcoupform}) 
has been found using Brauer algebras which have been 
useful in diagonalisation problems   of the CFT-metric on 
gauge invariant Matrix operators in four dimensional  $N=4$ SYM gauge theory. 
 These diagonalisation problems   
 have been useful \cite{cjr}  in mapping  gauge theory states to AdS-spacetime
 states such as $3$-brane configurations
 (giant gravitons). Brauer algebras arise in the case where we have both 
 branes and anti-branes \cite{kimram}.
 The map $ \Sigma $ which has been crucial  in developing
 formulae for projectors in the Brauer algebra \cite{kimram},
 is also  used to map projectors in $\C ( S_{ m + n } ) $ to Brauer elements
 in \cite{bcm}. The appearance of the same algebraic structures 
in  describing  strings in the string theory dual of 2DYM and 
3-branes in the string theory dual of 4DSYM 
suggests that  the geometrical lessons of 2DYM will also 
 have consequences for 4DSYM.

We  expect that the reformulation of  the complete (coupled) expansion 
in terms of the  chiral theory can lead to a deeper mathematical 
understanding of the  large $N$ expansion of 2DYM theory. 
We venture  some speculative ideas along these lines. 
From a physical perspective we  
want to understand, in generality, the relation between  2DYM 
for $ \Sigma ( G ) $ and Matrix models such as the Kontsevich 
Matrix model  \cite{konts} which can exhibit the  relation with the 
geometry of the compactified moduli space of worldsheet 
complex structures $\overline { \cM}_{g,n}  $. 
There are some  early attempts in this direction \cite{ksw}. 
Certain discrete counting problems 
related to Hurwitz spaces for sphere target, some of which were 
considered in the context of 2D Yang Mills \cite{crestay},  
have been mapped to integrals over  the compactified moduli spaces
of complex structures  $ \overline { \cM_{g,n} } $  \cite{elsv}   and  
results from 2D gravity  \cite{wit2D}  have allowed explicit computations.
In fact it is known the chiral 2DYM computes an Euler character of 
Hurwitz spaces \cite{cmri}.  It should be possible to express
this Euler character in terms of integrals 
of cohomology classes over  $ \overline{\cM}_{g,n}$.
This should  lead to a better understanding of how to express 
the observables inserted in (\ref{chiralGinsert}), which give the 
complete ${1\over N} $ expansion of 2DYM,  in terms of  
classes on Hurwitz spaces  and in turn on $ \overline{\cM}_{g,n}$. 
In line with other recent mathematical 
developments related to $A$-model topological strings \cite{okpan,vakil},
the construction of the appropriate classes on Hurwitz space and 
$ \overline{\cM}_{g,n}$ should probably proceed by first 
introducing the more complicated 
compactification of stable maps  for Gromov-Witten theory
with $ \Sigma ( G )$ target space,  and then introducing virtual classes 
whose integrals are simpler. In such a scenario, the two moduli spaces
of equal Euler character described in section 4, 
have to be understood as  the localization loci of virtual classes 
in the stable compactification.  Research on these avenues  
would lead to new connections  between integrable hierachies and 
the large $N$ expansion of 2DYM. For  the case of sphere target, 
the  topological $\sigma$-models   are  already known to
be related  to a matrix model and integrable hierarchies \cite{ehy}, 
with  conjectural relations for more general target \cite{ehs}.  
An extension of analogous results to define a  Matrix Model 
of Euler characters of holomorphic maps related to  the balanced  topological strings
\cite{cmri,dijmoo}  (and supplemented with the worldheet versions of 
the observable (\ref{chiralGinsert}) )  which generate the ${1\over N} $ expansion of
2DYM would set the stage for a quantitative  understanding of 
these strings for more general target  spaces.

\vskip.3in 

{ \Large 
{ \centerline { \bf Acknowledgements }  } } 

\vskip.2in 
We thank  Robert de Mello Koch, Gregory Moore, Sunil Mukhi, Rodolfo Rosso
 for useful discussions/correspondence. SR is 
supported by a  STFC Advanced Fellowship and in part by the
EC Marie Curie Research Training Network MRTN-CT-2004-512194. 
YK is supported by STFC grant PP/D507323/1 
``String theory, gauge theory and gravity''.

%%%%%%%%%%%%%%%%%%%%%%%%%%%%%%%%%%%%%%%%%%%

\vskip1in  

\appendix 

\section{Appendix :  Useful Formulae}

The projector used in section 2 is 
\begin{eqnarray}
p_{R}=\frac{d_{R}}{m!}\sum_{\sigma}\chi_{R}(\sigma)\sigma
\end{eqnarray}
The appearance of $\Pi_1$ in  large $N$ expansions of 2DYM stems from 
\begin{eqnarray}
\left(\frac{m!}{d_{R}}\right)^{2}
=\sum_{s,t\in S_{m}}\frac{\chi_{R}(sts^{-1}t^{-1})}{d_{R}}
\end{eqnarray}
The delta function used extensively in 2DYM  has   a character expansion 
\begin{eqnarray}
\frac{1}{n!}\sum_{R}d_{R}\chi_{R}(\rho)=\delta(\rho)
\end{eqnarray}
The relation between dimensions and the inverse $\Omega$  factor is  
\begin{eqnarray}
\frac{1}{dimR}=\frac{m!}{N^{m}}\frac{\chi_{R}(\Omega_{m}^{-1})}{d_{R}^{2}}
\end{eqnarray}
Useful formulae for manifolds with boundary are 
\begin{eqnarray}
&&
tr(\sigma U)=\sum_{R}\chi_{R}(\sigma)\chi_{R}(U) \cr
&&
\int dU \chi_{R}(U)\chi_{S}(U^{\dagger})=\delta_{RS}
\end{eqnarray}

%%%%%%%%%%%%%%%%%%%%%%%%%%%%%%%%%%%%

\section{ Appendix : Derivations  for  genus $G$  } 

For any decomposition of genus $G$ into pants and  one-holed tori we can 
write the full partition function in a way that reflects the choice of
 decomposition. 
Then going from chiral to full theory involves inserting the sum over 
subgroups 
and the projections of all the permutations using the symmetric projector
 of the subgroup.

We can write the genus $G$ answer from \cite{gt} 
\bea 
Z_{ G } &=&  \sum_{ m,n }  { N^{(m+n) (2-2G ) } \over m! n! }   \delta_{m,n} (
 \Omega_{m,n}^{2-2G } \Pi_1^G ) \cr  
&=&  \sum_{ m,n }  { N^{(m+n) (2-2G ) } \over m! n! } 
\sum_{ \alpha_1 \cdots \alpha_G \in S_m \times S_n } 
\delta_{m,n} ( (  \Omega_{m,n}^{-1} )^{G-2 }   \prod_i \alpha_i )  \prod_{i=1}^{G} 
  \delta_{m,n} (   \Omega_{m,n}^{-1}   \Pi_1  \alpha_i^{-1}  )  
\cr 
 &= & \sum_{ m,n }  { N^{(m+n) (2-2G ) } \over m! n! } ~ \sum_{ \alpha_1 
\cdots \alpha_G \in S_m \times S_n } 
 ~~ \prod_{i=1}^{G}   \delta_{m,n} (   \Omega_{m,n}^{-1}   \Pi_1 \alpha_i^{-1}  ) 
 \cr 
&& \sum_{ \beta_1 \cdots \beta_{G-3 } \in S_m \times S_n  } \delta_{m,n} ( \Omega_{m,n}^{-1} \alpha_1 
\alpha_2 \beta_1 )  \delta_{m,n} ( \Omega_{m,n}^{-1} \beta_1^{-1}  \alpha_3 \beta_2 ) \cr 
&&  \hskip2.5in 
 \delta_{m,n} (  \Omega_{m,n}^{-1} \beta_2^{-1}  \alpha_4 \beta_3 ) \cdots \delta_{m,n} (
    \Omega_{m,n}^{-1} \beta_{G-3} \alpha_{G-1}    \alpha_G )    \cr 
&=& \sum_{ m,n }  { N^{(m+n) (2-2G ) } \over m! n! } ~ \sum_{ \alpha_1 \cdots
 \alpha_G \in S_m \times S_n } 
 ~~ \prod_{i=1}^{G}   \delta_{m+n}  (   \Omega_{m+n}^{-1}  
 \Pi_1 \alpha_i^{-1}  )  \cr 
 && \sum_{ \beta_1 \cdots \beta_{G-3}  \in S_m \times S_n }
 \delta_{m+n} ( \Omega_{m+n}^{-1} \alpha_1 \alpha_2 \beta_1 )  \delta_{m+n}
 ( \Omega_{m+n}^{-1} \beta_1^{-1}  \alpha_3 \beta_2 ) \cr 
&& \hskip2.5in  \delta_{m+n}  (  \Omega_{m+n}^{-1} \beta_2^{-1}  \alpha_4 \beta_3 )
 \cdots \delta_{m+n} (    \Omega_{m+ n}^{-1} \beta_{G-3} \alpha_{G-1}   
 \alpha_G )   \nn \\ 
\eea 
After manipulating so that each delta functions contains a simgle 
power of $ \Omega^{-1} $ we can write in terms of 
the $ \Omega_{m+n}^{-1} $ leaving the $ \delta$ to do the projection.

We can also re-write in terms of any decomposition of the genus $G$  into 3-holed spheres, by imitating 
steps analogous to (\ref{gen2-3holeglue}) 
\bea
Z_{ G } &=&
\sum_{ m , n }  {   N^{ ( 2-2G ) ( m+n) }\over  m! n! }  \sum_{ s_1 , t_1 \cdots s_G , t_G  \in S_m \times S_n }     \delta_{m,n} ( \Omega_{m,n}^{2-2G } ~ s_1t_1 s_1^{-1} t_1^{-1} ~ s_2t_2 s_2^{-1} t_2^{-1}  \cdots 
                            s_G t_G s_G^{-} t_{G}^{-1} ) \cr 
&=&  \sum_{ m , n }  {   N^{ ( 2-2G ) ( m+n) }\over 
 m! n! }  \sum_{ s_1 , t_1 \cdots s_{G+1},  t_{G+1}  \in S_m \times S_n } 
  \delta_{m,n}  ( \Omega_{m,n}^{1-G } ~ s_1 s_2 \cdots  s_G ~ s_{G+1} ) \cr 
&& \hskip1in  \delta_{m,n} ( \Omega_{m,n}^{1-G } ~ t_1 s_1^{-1} t_1^{-1} ~ t_2 s_2^{-1} t_2^{-1} \cdots t_G s_G^{-1} t_G^{-1}  s_{G+1}^{-1} ) \cr 
&=& \sum_{ m , n } { N^{(2-2G)(m+n) }      \over (m! n!)^2  } \sum_{s_i , t_i, u_i \in S_m \times S_n   } 
\delta_{m,n} (  \Omega_{m,n}^{1-G } ~ s_1 s_2 \cdots  s_G ~ s_{G+1} ) 
 ~~  \delta_{m,n} ( \Omega_{m,n}^{1-G } u_1 u_2 \cdots u_G u_{G+1} ) \cr 
&& \hskip1in ~~ \prod_{i=1}^{G+1} \delta_{m,n} ( u_i t_i s_i^{-1} t_i^{-1} ) \cr 
&=& \sum_{ m , n } { N^{(2-2G)(m+n) }      \over (m! n!)^2  } \sum_{s_i,t_i,u_i} \sum_{ v_1, w_1 \cdots v_{G-2} , w_{G-2}  \in S_m \times S_n } 
 \delta_{m,n} ( \Omega_{m,n}^{-1} s_1 s_2 v_1 ) ~~ \delta_{m,n} ( \Omega_{m,n}^{-1} u_1 u_2 w_1 ) \cr 
 &&\hskip1in  \delta_{m,n} ( \Omega_{m,n}^{-1} v_1^{-1}  s_3 v_2 ) ~~ \delta_{m,n} ( \Omega_{m,n}^{-1} w_1^{-1}  u_3 w_2 ) \cr 
&& \hskip1in \qquad \hskip.5in  \vdots  \hskip1in \qquad \vdots \cr 
&& \hskip1in \delta_{m,n} ( \Omega_{m,n}^{-1}  v_{G-3}^{-1} s_{G-1} v_{G-2} )  \delta_{m,n} ( \Omega_{m,n}^{-1}  w_{G-3}^{-1} s_{G-1} w_{G-2} ) \cr 
&& \hskip1in \delta_{m,n} ( \Omega_{m,n}^{-1} v_{G-2}^{-1} s_G s_{G+1} ) 
 \delta_{m,n} ( \Omega_{m,n}^{-1} w_{G-2}^{-1} u_G u_{G+1} ) \cr 
&& \hskip1in \prod_{i=1}^{G+1} \delta_{m,n} ( u_i t_i s_i^{-1} t_i^{-1} ) \cr 
 &=& \sum_{ m , n } \sum_{s_i,t_i,u_i, v_i , w_i ,\tv_i ,\tw_i , \g_i , \e_i  \in S_m \times S_n  }
  { N^{-(m+n) }      \over (m! n!)  } \delta_{m,n} ( \Omega_{m,n}^{-1} s_1 s_2 v_1 ) ~~ { N^{-(m+n) }      \over (m! n!)  } 
\delta_{m,n} ( \Omega_{m,n}^{-1} u_1 u_2 w_1 ) \cr 
 &&  \hskip1in { N^{-(m+n) }      \over (m! n!)  } \delta_{m,n} ( \Omega_{m,n}^{-1}  \tv_1^{-1}  s_3 v_2 ) ~~  { N^{-(m+n) }      \over (m! n!)  } \delta_{m,n} ( \Omega_{m,n}^{-1} \tw_1^{-1}  u_3 w_2 ) \cr 
&& \qquad  \hskip3in  \vdots \hskip1.5in \vdots \cr 
&&  \hskip1in  { N^{-(m+n) }      \over (m! n!)  } \delta_{m,n} ( \Omega_{m,n}^{-1}  \tv_{G-3}^{-1} s_{G-1} v_{G-2} )
   { N^{-(m+n) }      \over (m! n!)  }
\delta_{m,n} ( \Omega^{-1}  \tw_{G-3}^{-1} s_{G-1} w_{G-2} ) \cr 
&&  \hskip1in  { N^{-(m+n) }      \over (m! n!)  } \delta_{m,n} ( \Omega_{m,n}^{-1} 
\tv_{G-2}^{-1} s_G s_{G+1} )  { N^{-(m+n) }      \over (m! n!)  }  \delta_{m,n}
 ( \Omega_{m,n}^{-1} \tw_{G-2}^{-1} u_G u_{G+1} ) \cr 
&&  \hskip1in \prod_{i=1}^{G+1} \delta_{m,n} ( u_i t_i s_i^{-1} t_i^{-1} )    
\prod_{ i=1}^{G-2}  \delta_{m,n} ( v_i^{-1}  \g_i  \tv_i   \g_i^{-1} ) 
                          \prod_{ i=1}^{G-2}  \delta_{m,n} ( w_i^{-1}  \e_i  \tw_i 
  \e_i^{-1} )         
\eea 

Now that all the $\Omega^{-1} $  factors are sitting 
in separate delta functions in $ S_m  \times S_n$ 
along with permutations within that subgroup, we may obtain 
(\ref{genGpantseq}).

\section{Appendix : Omega factors}
\label{sec:omegachiralandcoupled}
In this section, we show explicit forms of $\Omega_{m+n}^{-1}$ and
$\Omega_{m,n}^{-1}$ 
for some examples. 
One way to calculate 
$\Omega_{m+n}^{-1}$ is to solve 
$\Omega_{m+n}\Omega_{m+n}^{-1}=1$. 
Another useful way is to use 
\bea 
\Omega_{m+n}^{-1} = \frac{N^{m+n}}{((m+n)!)^{2}} 
\sum_{ T \vdash (m+n) }
 { d_T^2 \over Dim T } \chi_{T} ( \sigma ) \sigma 
\eea 
which was used in  \cite{kimram} to obtain the dual of 
Brauer algebra elements with respect to a bilinear form. 

When $m+n=3$, the omega factor is given by 
\begin{eqnarray}
&&
\Omega_{3}=1+\frac{1}{N}T_{[2,1]}+\frac{1}{N^{2}}T_{[3]}
\end{eqnarray}
The indices written as subscripts of $T$ denote the cycle lengths
of the conjugacy class.  
The inverse of this is calculated using the above formula as
\begin{eqnarray}
\Omega_{3}^{-1}=\frac{N^{2}}{(N^{2}-1)(N^{2}-4)}
\left(
N^{2}-2-NT_{[2,1]}+2T_{[3]}
\right)
\end{eqnarray}
By projecting this to the subgroup 
$S_{2}\times S_{1}$, we get
\begin{eqnarray}
\Omega_{3}^{-1}|_{S_{2}\times S_{1}}
=\Omega_{2,1}^{-1}=
\frac{N^{2}}{(N^{2}-1)(N^{2}-4)}
\left(
N^{2}-2-Ns_{1}
\right)
\end{eqnarray}
We can easily check $\Omega_{2,1}^{-1}\Omega_{2,1}=1$ 
using 
\begin{eqnarray}
&&
\Omega_{2,1}=1-\frac{2}{N^{2}}+\frac{1}{N}s_{1}
\end{eqnarray}

Another example is the case of $m+n=4$, where the inverse of the omega
factor is  
\begin{eqnarray}
\Omega_{4}^{-1}
&=&
\frac{N^{2}}{(N^{2}-1)(N^{2}-4)(N^{2}-9)}\times  \cr
&&
\Big(N^{4}-8N^{2}+6-N(N^{2}-4)T_{[2,1^{2}]}
+(2N^{2}-3)T_{[3,1]} \cr
&&
-5NT_{[4]}+(N^{2}+6)T_{[2,2]}
\Big) 
\end{eqnarray}
In this case, 
we can consider two projections to $S_{3}\times S_{1}$ and 
$S_{2}\times S_{2}$, which give  
\begin{eqnarray}
\Omega_{4}^{-1}|_{S_{3}\times S_{1}}
=\Omega_{3,1}^{-1}
&=&
\frac{N^{2}}{(N^{2}-1)(N^{2}-4)(N^{2}-9)}\times  \cr
&&
\Big(N^{4}-8N^{2}+6-N(N^{2}-4)T_{[2,1]}
+(2N^{2}-3)T_{[3]} 
\Big) 
\end{eqnarray}
and 
\begin{eqnarray}
\Omega_{4}^{-1}|_{S_{2}\times S_{2}}
=\Omega_{2,2}^{-1}&=&
\frac{N^{2}}{(N^{2}-1)(N^{2}-4)(N^{2}-9)}\times \cr
&&
\Big(N^{4}-8N^{2}+6-N(N^{2}-4)(s+\bar{s})
+(N^{2}+6)s\bar{s}
\Big) 
\end{eqnarray}
These can also be checked using 
\begin{eqnarray}
\Omega_{3,1}
&=&
1-\frac{3}{N^{2}}
+\frac{1}{N}
\left(1-\frac{1}{N^{2}}\right)
T_{[2,1]}
+\frac{1}{N^{2}}T_{[3]}
\end{eqnarray}
and
\begin{eqnarray}
\Omega_{2,2}
&=&
1-\frac{4}{N^{2}}+\frac{2}{N^{4}}
+\frac{1}{N}(s+\bar{s})
+\frac{1}{N^{2}}\left(1-\frac{2}{N^{2}}\right)s\bar{s}
\end{eqnarray}
Here $s$ is the transposition in the left $S_2 $ factor, 
while $ \bar s $ is the transposition in the right $S_2$ 
factor.

 According to (\ref{newcoupform}) and the above explicit 
 formulae,   all  the restrictions on the nature of the double 
 points in $ \Omega_{m,n}$
  reviewed in section 4.1  are encoded in the inversion of  
 $ \Omega_{m+n}$ to give $ \Omega_{m+n}^{-1} $. 
 This uses the properties
 of group multiplication  inside $S_{m+n}$ 
 followed by projection to $S_m \times S_n$. It is not a 
 straightforward relation such as saying that the double points 
 of the coupled expansion follow from all those arising in the 
 collision of branch points as encoded in symmetric group 
 multiplication. For example general products of permutations can give nodes 
 connecting different types of cycles. Consider multiplications 
 in $S_4$ such as   $(132)(1234) = (1)(2) (34) $. The counting of branching 
 numbers implies that after such a collision there are two nodes degenerated 
 at one point. Such nodes do not arise the coupled expansion. 
 The above multiplication does however enter the relation between 
 $ \Omega_{4}^{-1} $ and $\Omega_{2,2}^{-1} $. 
 If the reduction 
is to $ S_{ \{ 1,2\} }  \times S_{\{ 3 , 4 \} }  $ then it seems
 we have tubes connecting 
2-cycles to 1-cycles. This  not so.
We can think of $(1)(2) (34) $ as a product of 
trivial permutation with the $(1)(2) (34) $. We associate ${ 1 \over N^2 } $ 
with the trivial  permutation. In other words in expanding 
$ \Omega_{m,n}^{-1} = ( 1 + \tilde \Omega_{m,n} )^{-1}  $ 
the term corresponding to $N^{-5} (132)(1234) $ in the expansion of 
 $\Omega_{4}^{-1} $ 
comes from  the $ ( \tilde \Omega_{m,n} )^2 $ and not from
 $  \tilde \Omega_{m,n}$.

\section{Appendix: Gluing manifolds with boundary}

Gauge theory partition functions are defined as a function 
of the boundary holonomy $U$ which lives in the gauge group $U(N)$.
 They can be calculated exactly.  
Consider, for example,  the case $ \Sigma ( G =1 , B=1 ) $.  
\begin{eqnarray}
Z(G=1 , B=1 ; U)=\sum_{R}\frac{1}{dimR}\chi_{R}(U)
\end{eqnarray}
To get observables appropriate for a string interpretation in the chiral expansion we 
 choose  a positive integer $M$
and a conjugacy class $T$ in $S_M$.  
\begin{eqnarray}
Z^+ ( G=1 , B=1 ; T )
&\equiv & \int dU Z^+(U)~\frac{1}{M!}\sum_{ \alpha \in T }  \sum_{R } 
\chi_R ( \a ) \chi_R ( U^{\dagger}  )  \cr
&=&
\frac{N^{-M}}{M!}
\sum_{s,t\in S_{M}}\sum_{ \alpha \in T }
\delta_M (\Omega_{M}^{-1} sts^{-1}t^{-1} \a)
\end{eqnarray}
The delta function is defined over the group algebra 
of $S_M$.  
This can be interpreted as a sum 
over covering spaces of $ \Sigma ( G =1 , B=1 )$ 
subject to the constraints that the permutation of
 the sheets upon going on a path round the boundary is 
in the conjugacy class $T$. 
After expanding the $ \Omega $ factors as in \cite{cmri} 
\bea 
&& Z^+ (G=1 , B=1 ;  T )=
\frac{N^{-M}}{M!}\sum_{ L=0}^{\infty} 
\sum_{\sigma_{1},\cdots,\sigma_{L}\in S_{M}}^{\prime } 
\sum_{s,t\in S_{M}}\sum_{ \alpha \in T }
(-1)^{L}  \cr 
&& \hskip2in  \left(\frac{1}{N}\right)^{\sum_{j=1}^{L}(M-C_{\sigma_{j}})}
 \delta_M ( \sigma_1 \cdots \sigma_L  sts^{-1}t^{-1} \a)  
\eea
The factor $(-1)^L$ is the Euler character of 
the $L$-dimensional  configuration space of $L$ branch points on 
$ \Sigma( G=1 , B=1 )$, hence the interpretation as an Euler 
character of the moduli space of branched covers. This 
is explained in detail in \cite{cmri,cmr}

In the non-chiral theory, we choose two integers $ m,n$ and
  a conjugacy class $T$  in $S_m \times S_n$. 
and we  multiply $ Z(U)$  with characters 
$  \sum_{ \sigma \in T  }  \chi_{ R \otimes S } ( \sigma ) 
  \chi_{R \bar S } ( U^{\dagger}  )$. 
\bea 
Z (G=1, B=1;  T ) &=&\sum_{ \alpha \in T }  Z ( G=1 , B=1 ; \alpha ) \cr 
&=&
\sum_{ \a \in T } \frac{N^{-m-n }}{m!n!}\sum_{s,t\in S_{m} \times S_n }
\delta_{m,n}  ( \Omega_{m,n}^{-1} sts^{-1}t^{-1} \alpha)
\eea
The delta function is defined over the group algebra of $S_m \times S_n$. 
Given the formula for $ \Omega_{m,n} $ in (\ref{coupomegform})
this can be interpreted in terms branched coverings 
from   wordsheets made of pairs
of surfaces joined at double points. One component of the pair 
maps holomorphically, the other maps anti-holomprphically. 
After expanding the $ \Omega_{m,n} $ factor we can interpret 
$ Z( G=1, B=1; T ) $ as an Euler character of a moduli space of
``coupled maps.'' By using an orientation reversal, the coupled maps 
are nothing but degenerated holomorphic maps.

The gluing of partition functions is done by 
integrating over the $U(N)$ holonomy $U$ along a common boundary. 
This can be translated into  a rule for how to glue the 
partition functions with boundary data in terms of symmetric groups. 
In the chiral theory, the rule is to sum over all possible $T$ for all $M$. 
We describe the gluing of two copies of $ \Sigma ( G=1 , B=1)$ 
to get $ \Sigma ( G=2 , B=0 )$ 
\bea 
&& Z^+   ( G = 2 ) = \sum_{  M } 
 \sum_{ \alpha_1, \b_1    \in S_M  }  
                  Z^+ ( G=1 , \a_1   )  Z^{+} ( G=1 , \b_1^{-1}  )
  \cr 
&& \hskip2.2in \sum_{ \g \in S_M  } 
 \delta_M  ( \alpha_1^{-1}  \gamma \b_1 \gamma^{-1} ) 
\eea 
We expect the gluing of boundary partition functions 
for two copies of $ Z_{ G=0, B=3 } $ to give $ Z( G=2) $ 
and indeed the following equality holds 
\bea 
&& Z^{+}  ( G=2 ) = \sum_{ M } \sum_{ \alpha_i , \beta_i \in S_M  } 
Z^+ ( G=0 , B=3 ;  \alpha_1 , \alpha_2 , \alpha_3  )
Z^+ ( G=0 , B=3 ; \beta_1^{-1}  , \beta_2^{-1}  , \beta_3^{-1}  )   \cr  
&&   \hskip2in       \prod_{ i=1}^{3} \sum_{ \gamma_i }
        \delta_M  ( \alpha_i^{-1}  \gamma_i \beta_i \g_i^{-1} )
\eea 

Analogous results for the non-chiral partition function 
are 
 \bea 
&& Z   ( G = 2 ) = \sum_{m, n } \quad 
 \sum_{ \alpha_1, \b_1   \in S_m \times S_n   }  
                  Z ( G=1 , \a_1   )  Z ( G=1 , \b_1^{-1}  )
  \cr 
&& \hskip2.2in \sum_{ \g \in  S_m \times S_n   } 
 \delta_{m,n}  ( \alpha_1^{-1}  \gamma \b_1 \gamma^{-1} ) 
\eea   

and 
\bea 
&& Z  ( G=2 ) = \sum_{ m,n } 
\sum_{ \alpha_i , \beta_i \in S_m \times S_n }  
Z ( G=0 , B=3 ;  \alpha_1 , \alpha_2 , \alpha_3  )
Z ( G=0 , B=3 ; \beta_1^{-1}  , \beta_2^{-1}  , \beta_3^{-1}  )   \cr  
&&   \hskip2in       \prod_{ i=1}^{3} \sum_{ \gamma_i\in S_{m}\times S_{n} }
        \delta_{m,n}  ( \alpha_i^{-1}  \gamma_i \beta_i \g_i^{-1} )
\eea

%%%%%%%%%%%%%%%%%%%%%%%%%%%%%%%%%

\end{document}